# Resilient Growth of Highly Crystalline Topological Insulator-Superconductor Heterostructure Enabled by Ex-situ Nitride Film

*Renjie Xie,[1,#] Min Ge,[2,#] Shaozhu Xiao,[3] Jiahui Zhang,[1] Jiachang Bi,[1] Xiaoyu Yuan,[4] Hee Taek Yi,[4] Baomin Wang,[5] Seongshik Oh,[4] Yanwei Cao,[1,*] and Xiong Yao[1,*]*

[1]Ningbo Institute of Materials Technology and Engineering, Chinese Academy of Sciences, Ningbo 315201, China

[2] The Instruments Center for Physical Science, University of Science and Technology of China, Hefei 230026, China

[3] Yongjiang Laboratory, Ningbo 315202, China

[4]Department of Physics & Astronomy, Rutgers, The State University of New Jersey, Piscataway, New Jersey 08854, United States

[5] School of Physical Science and Technology, Ningbo University, Ningbo 315211, China

# Renjie Xie and Min Ge contribute equally to this work

*Email: ywcao@nimte.ac.cn

*Email: yaoxiong@nimte.ac.cn

ABSTRACT

Highly crystalline and easily feasible topological insulator-superconductor (TI-SC) heterostructures are crucial for the development of practical topological qubit devices. The optimal superconducting layer for TI-SC heterostructures should be highly resilient against external contaminations and structurally compatible with TIs. In this study, we provide a solution to this challenge by showcasing the growth of a highly crystalline TI-SC heterostructure using refractory TiN (111) as the superconducting layer. This approach can eliminate the need for in-situ cleaving or growth. More importantly, the TiN surface shows high resilience against contaminations during air exposure, as demonstrated by the successful recyclable growth of $Bi_2Se_3$. Our findings indicate that TI-SC heterostructures based on nitride films are compatible with device fabrication techniques, paving a path to the realization of practical topological qubit devices in the future.

## INTRODUCTION

Majorana fermion modes (MFMs), as a type of excitation that obeys non-Abelian statistics, are widely regarded as a promising platform for implementing topological qubits in fault-tolerant quantum computation. Several approaches have been proposed to explore MFMs in condensed matter systems. The zero-dimensional manifestation of MFMs, known as Majorana bound states, has been theoretically predicted to exist at the vortices of an s-wave superconductor/topological insulator heterointerface[1, 2], and indeed, the signature of Majorana modes has been experimentally detected within the vortex core of a $Bi_2Te_3$-$NbSe_2$ heterostructure[3, 4]. The main challenge along this approach is to exclude the contribution from other trivial effects such as disorder effect, Kondo correlations, Andreev-bound states, etc. Another scheme to create one-dimensional MFMs involves coupling a superconductor with a quantum anomalous Hall insulator, which can be achieved in a superconductor-magnetic topological insulator heterostrcuture[5, 6]. This approach not only provides signatures of MFMs that can be directly detected by transport measurements, but also allows for the readout of topological qubits. Nonetheless, the existence of MFMs in the proposed device setup still remains controversial due to the lack of ideal heterostructures that can meet all the proposed reqirements[7, 8]. Evidently, the fabrication of topological insulator-superconductor (TI-SC) heterostructure is pivotal in various proposals aimed at MFMs, emphasizing the importance of highly crystalline and easily feasible TI-SC heterostructures.

Various TI-SC heterostructures have been fabricated so far in the search for MFMs[3, 4, 7-20]. The obstacles to reaching an easily feasible TI-SC heterostructure mainly arise from two aspects. First, owing to the reactive nature of the superconducting layers, these heterostructures are either grown on in-situ cleaved surface of superconducting bulk crystals[3, 4, 9, 11, 14] or in-situ grown superconducting films[12, 15-17], to avoid degradation of the interface. The in-situ cleaving and growing process not only impedes the feasibility of TI-SC heterostructures but is also incompatible with device fabrication processes such as lithography and etching, creating barriers for the development of practical topological qubit devices. Second, considering the structural match at the interface has been proven to be critical for improving the sample quality[21-23], the mismatched structural symmetry between TI and iron-based or cuprate superconductors (6-fold vs 4-fold) remains a big challenge to growing high-quality TI-SC heterostructures[20, 24-26]. The ideal superconducting layer for TI-SC heterostructure should be highly resilient against contaminations in air exposure and structurally compatible with TIs. Here we demonstrate an unprecedented, recyclable growth of a highly crystalline TI-SC heterostructure, achieved by utilizing ex-situ grown TiN (111) film that functions akin to such an ideal superconducting layer. The robust TiN surface that resists various forms of degradation, pollution, and thermal treatment as demonstrated here, is highly promising for the development of practical TI-SC heterostructures and devices.

## RESULTS AND DICUSSION

Figure 1a shows the schematic illustration of the heterostructure consisting of a 3D TI layer ($Bi_2Se_3$) and a superconductor layer (TiN). Evidently, the TiN (111) layer shares the same 6-fold in-plane structural symmetry with the 3D TI $Bi_2Se_3$[27, 28], facilitating the growth of high-quality heterostructure owing to the structural match. Epitaxial TiN (111) films with thickness around 70 nm were grown on 5 mm × 5 mm $Al_2O_3$ (0001) substrates by a homemade reactive magnetron sputtering system. Subsequently, the as-grown TiN films were taken out from the magnetron sputtering system and transferred into a homemade ultrahigh vacuum (UHV) molecular-beam epitaxy (MBE) chamber. The TiN film was annealed at 700 ℃ under a UHV environment for 12 h to clean the surface. The high quality of the TiN films was validated through the distinct, well-defined low-energy electron diffraction (LEED) patterns, as illustrated in Figure 1b. Figure 1c gives the sharp spotty LEED pattern of $Bi_2Se_3$, from which the in-plane lattice constant can be determined as 4.1 Å, consistent with the previously reported value[29].

The crystallinity of the $Bi_2Se_3$/TiN films was characterized by X-ray diffraction (XRD) scans, as shown in Figure 1d. All the $Bi_2Se_3$ (00 3l) peaks are clearly observable without any sign of a second phase. We compared the rocking curve of a $Bi_2Se_3$/TiN film with a $Bi_2Se_3$ film directly grown on $Al_2O_3$ under identical growth conditions, as seen in Figure 1e. The FWHM (Full Width at Half Maximum) values, extracted from the rocking curves, are 0.07° and 0.27° for $Bi_2Se_3$/TiN/ $Al_2O_3$ and $Bi_2Se_3$/$Al_2O_3$, respectively. Such a comparison strongly indicates that the crystallinity of $Bi_2Se_3$/TiN/$Al_2O_3$ even outperforms $Bi_2Se_3$/$Al_2O_3$. We further compared the atomic force microscopy (AFM) image of $Bi_2Se_3$/TiN/$Al_2O_3$ and $Bi_2Se_3$/$Al_2O_3$ in Figures 2a and b. The surface morphology of both samples shows characteristic triangular-shaped terraces with a height of around 1 nm (corresponding to 1 QL of $Bi_2Se_3$), while the terraces of $Bi_2Se_3$/TiN/$Al_2O_3$ are obviously larger than that of $Bi_2Se_3$/$Al_2O_3$. The large flat terraces of $Bi_2Se_3$/TiN/$Al_2O_3$ are direct evidence of the well-matched structural symmetry provided by the TiN surface[22]. Notably, as exhibited in Figure S1a and b, the surface flatness of TiN film is even better than commercial $Al_2O_3$ substrate, suggesting the obvious advantage of TiN for TI-SC heterostructures. Considering that our $Bi_2Se_3$/TiN heterostructures are grown on ex-situ grown TiN films and have not involved any in-situ cleaving or growing process, the high crystallinity and large terraces unveil the huge potentials of utilizing TiN film as the superconducting layer in practical TI-SC heterostructures.

To further investigate the interface structure of the $Bi_2Se_3$/TiN heterostructure at the atomic scale, we performed the scanning transmission electron microscope (STEM) characterizations. Figure 2c presents the cross-sectional high-angle annular dark-field (HAADF)-STEM imaging acquired at the interface between $Bi_2Se_3$ and TiN. The interface is atomically sharp and well-defined flat over several tens of nanometers. The $Bi_2Se_3$ layer is highly crystalline as demonstrated by the uniform van der Waals gap without any atomic distortion or interlayer diffusion. The corresponding energy-dispersive X-ray (EDX) mappings in Figure 2c reaffirm the atomically smooth and sharp interface without any atomic interdiffusion across the interface, which is critical for realizing the desired superconducting proximity effect in TI-SC heterostructures. Figures 2d and e give the Fourier-transformed images at the $Bi_2Se_3$ layer and TiN layer, respectively. Clear patterns can be seen in both $Bi_2Se_3$ and TiN regions. The Fourier-transformed images suggest the epitaxial growth of $Bi_2Se_3$ (00 3l) along TiN (111) in the c direction, consistent with the XRD results.

The superconducting properties of $Bi_2Se_3$/TiN were investigated by transport measurements using a Quantum Design physical properties measurement system (PPMS). In Figure 3a, the temperature-dependent longitudinal resistance of a $Bi_2Se_3$/TiN heterostructure shows a clear and sharp superconducting transition at 5.4 K, implying the superconductivity of TiN remains intact after the deposition of $Bi_2Se_3$ layer. Figure 3b gives the temperature-dependent longitudinal resistance measured under varying magnetic fields perpendicular to the ab plane. The superconducting transition becomes broader and gradually shifts to lower temperatures under increasing applied magnetic fields.

To confirm that the pristine topological band structure is preserved at the $Bi_2Se_3$ surface, we performed angle-resolved photoemission spectroscopy (ARPES) measurement at the top surface of a 15 nm $Bi_2Se_3$ grown on TiN, the result is shown in Figure 3c. The ARPES pattern clearly shows the V-shaped Dirac surface band, implying the non-trivial topology of the $Bi_2Se_3$ layer. The energy difference between the Dirac point and Fermi level is determined as 0.22 eV in Figure 3c, obviously lower than the value of 0.33 eV in low-disorder $Bi_2Se_3$ grown on $Al_2O_3$ and only slightly higher than the value of 0.17 eV in $Bi_2Se_3$ grown on buffer layer template[30, 31]. Our ARPES result suggests the Fermi level of our $Bi_2Se_3$/TiN heterostructure is closer to the Dirac point compared with $Bi_2Se_3$/$Al_2O_3$, either due to the reduction of interfacial defects facilitated by the improved structural match and surface flatness of TiN or as a result of charge transfer from the TiN layer. In either case, this observation reveals another advantage of $Bi_2Se_3$/TiN heterostructure since it is easier to achieve intrinsic topological insulating $Bi_2Se_3$ on TiN than on other substrates.

The above results in Figures 1-3 show that ex-situ growth of highly crystalline TI-SC heterostructure can be realized in $Bi_2Se_3$/TiN, eliminating the complex and strict procedures required for in-situ cleaving or in-situ fabrication. Next, we demonstrate the exceptionally resilient TiN surface against undesirable pollutions or degradations in air exposure by showcasing a recyclable growth of $Bi_2Se_3$/TiN, which was previously inaccessible in any other TI-SC heterostructures. Figure 4 presents the detailed procedures of the recyclable growth together with the corresponding sample color and LEED pattern at each step. Here all the pictures of the sample were taken in situ. We started with a fresh annealed TiN film, which shows the characteristic golden color of TiN as well as sharp LEED spots in Figure 4a. We followed the same two-step protocol described previously to deposition 30 nm $Bi_2Se_3$ on top of TiN. The sample color changed to uniform metal color and the LEED pattern transitions to $Bi_2Se_3$ in Figure 4b. Then we took the sample out of the MBE chamber and performed ex-situ AFM measurements, as shown in Figure 4c. The AFM topography shows the characteristic morphology of $Bi_2Se_3$ in ordered triagonal-shaped terraces. The size of the terraces is smaller than Figure 2b due to the thinner thickness of $Bi_2Se_3$. Figure 4c shows that various undesired pollutions such as $H_2O$, $O_2$, H+, etc. are absorbed to the sample and lead to the degraded surface. We transferred this degraded sample back into the MBE chamber and recorded the sample status in Figure 4d. The sample color remains unchanged, and the exposed golden region (TiN) at the corner is due to the slight misalignment of sample holder positions during two separate sample mountings. The LEED spots become very dim and blurry, confirming the surface degradation during air exposure. Then we heated up the sample to 700 ℃ for 12h to evaporate the $Bi_2Se_3$ layer together with the polluted surface on top. The sample color recovered to the shiny golden color of TiN, and the LEED spots changed to a sharp and bright TiN pattern in Figure 4e. We repeated the deposition of $Bi_2Se_3$ in Figure 4f to test whether the recovered TiN surface can provide the same ideal flatness and interfacial match for TI growth. After the $Bi_2Se_3$ deposition, the color changes back to metal color, and the LEED pattern is similarly sharp and bright with the first time grown $Bi_2Se_3$ in Figure 4b.

We took out this sample and performed AFM and XRD measurements, as shown in Figure 4f. The AFM morphology exhibits terraces with smaller sizes and higher density compared with Figure 4c, while the XRD rocking curves show almost the same FWHM value as the first growth. Moreover, the surface morphology of the TiN layer became rougher after $Bi_2Se_3$ evaporation, as revealed by the AFM results shown in Figure S1c. These observations indicate the following scenario: Some residual atoms from the $Bi_2Se_3$ layer stick on the TiN surface after the evaporation and serve as nuclei for the second time $Bi_2Se_3$ growth, leading to a rougher surface and higher nucleation density compared with the fresh TiN surface. However, despite the increased nucleation density, the growth quality is controlled by the structural match of the TiN surface, which remains as good as fresh TiN. So, the increased nucleation density and structurally matched surface led to the AFM and XRD results in Figure 4f. More importantly, we compared the superconducting property of $Bi_2Se_3$/TiN heterostructure and TiN after $Bi_2Se_3$ evaporation. As shown in Figure S6, the resistance curves clearly overlap, confirming the stability and resilience of TiN against air exposure.

## CONCLUSIONS

The recyclable growth we demonstrated is not only for reusing TiN films in TI-SC heterostructures but more importantly, to showcase the resilient TiN surface against various external contaminations and deleterious effects. Such ability is unique among various superconducting films and critical for fabricating TI-SC heterostructure devices, which involve processes such as lithography and etching. For example, our results could remove the barriers toward TI-SC Josephson junctions and other patterned TI-SC devices[32-34]. The highly crystalline and easily feasible $Bi_2Se_3$/TiN heterostructure, along with its compatibility with device fabrication techniques, confirms TiN as the optimal superconducting layer for constructing TI-SC heterostructures. Compared to mechanical exfoliation strategy, the $Bi_2Se_3$/TiN heterostructure demonstrated here is highly reproducible, easily feasible and capable for device fabrication at millimeter scale, which is crucial for scaling up the TI-SC devices. Moreover, our results are applicable to other superconducting transition-metal nitride films (e.g., NbN, TaN, VN), which means there is a large category of nitride compounds that can serve as resilient and structurally compatible superconducting layers, paving the way toward practical TI-SC heterostructures for topological qubit devices in the future.

EXPERIMENTAL SECTION

**Preparation of $Bi_2Se_3$/TiN heterostructure:** Epitaxial TiN (111) films were grown on 5 mm × 5 mm $Al_2O_3$ (0001) substrates by a homemade reactive magnetron sputtering system, the details of which can be found in our previous reports[35-38]. Subsequently, the as-grown TiN films were taken out from the magnetron sputtering system and transferred into a homemade ultrahigh vacuum MBE chamber. The TiN film was annealed at 700 °C under a UHV environment for 12 h to clean the surface. $Bi_2Se_3$ thin films were grown on top of the TiN surface by thermally evaporating pure elemental Bi (99.997%) and Se (99.999%) sources using Knudsen cells. All the source fluxes were calibrated in situ by a quartz crystal microbalance. For $Bi_2Se_3$ growth, an initial 3 quintuple layers (QLs) seed layer was deposited at a relatively low temperature of 150°C, followed by the growth at 250°C.

Regarding the recyclable growth described in Figure 4, the first time $Bi_2Se_3$ deposition was conducted according to the same protocol described above. After growth, the sample was exposed to the atmospheric environment for several hours. It was then returned to the MBE chamber, annealed at 700°C for 12 hours to obtain a clean surface for the second time $Bi_2Se_3$ deposition. Then the second time $Bi_2Se_3$ deposition was repeated under the same growth conditions.

**Crystal structure and surface analysis**: The crystal structure was characterized by high-resolution X-ray diffraction (HRXRD, Bruker D8 Discovery) measurements, with Cu Kα radiation ($\lambda$ = 1.5406 Å) as the X-ray source. Symmetrical $\theta$- $2\theta$ measurements, rocking curves, X-ray reflectivity and $\varphi$ scans were performed with 40 kV working voltage and 40 mA current. LEED images were collected in-situ at the electron energy of 92 eV. The surface morphology was examined by a Bruker Dimension Icon AFM with Nanoscope V controller (Digital Instruments, USA). All topographic images were acquired in the scanAsyst mode with silicon cantilevers.

The scanning transmission electron microscopy (STEM) samples were prepared using focused ion beam (FIB) milling. High-angle annular dark-field scanning transmission (HAADF-STEM) and energy dispersive X-ray (EDX) spectroscopy measurements were conducted on the Themis Z Double spherical aberration corrected transmission electron microscope with 300 kV.

**Transport measurement:** The temperature-dependent resistance of TiN films and $Bi_2Se_3$/TiN heterostructure were measured using a Quantum Design physical properties measurement system (PPMS). Electrical electrodes were made by manually pressing four indium wires on each sample.

**Angle Resolved Photoemission Spectroscopy measurement:** The electronic structure of $Bi_2Se_3$/TiN heterostructure was measured by a home-made angle resolved photoemission spectroscopy (ARPES) with a 177 nm deep ultraviolet laser ($h_v$=6.997 eV). The base pressure was better than $5 \times 10^{-11}$ mbar. The total energy resolution was better than 1 meV. The sample was measured at about 7.3K. The Fermi level was determined by the measurement of a polycrystalline gold sample in electrical contact with the sample.

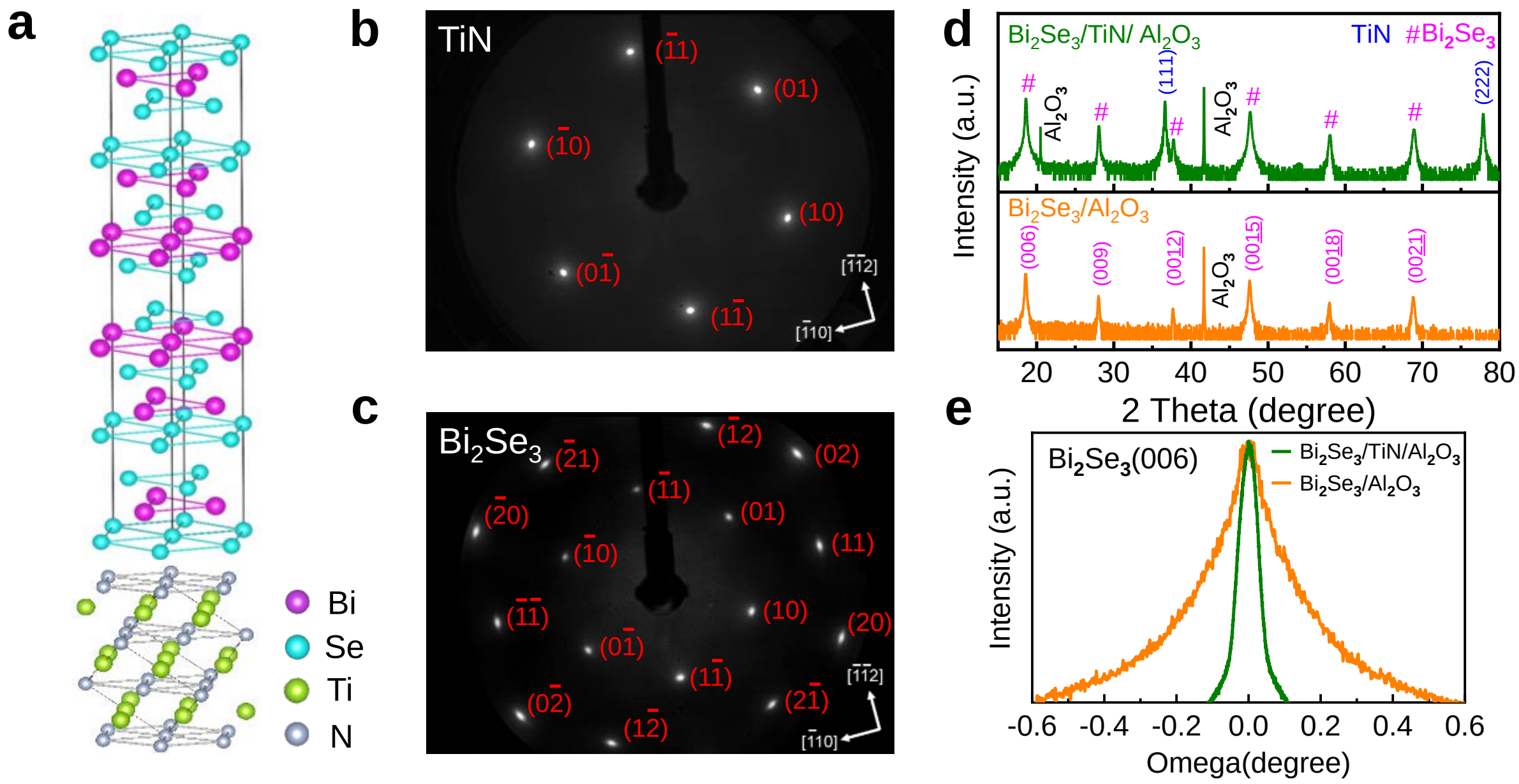

Figure 1. (a) Schematic illustration of the $Bi_2Se_3$/TiN heterostructure. (b, c) In situ LEED patterns of (b) TiN film and (c) $Bi_2Se_3$/TiN measured with electron energy of 92 eV. Integer order spots are marked. (d) The XRD profiles of $Bi_2Se_3$/TiN/$Al_2O_3$ and $Bi_2Se_3$/$Al_2O_3$. The marks labeled $Bi_2Se_3$ (00 3l) and TiN peaks. (e) Rocking curves of $Bi_2Se_3$/TiN/$Al_2O_3$ and $Bi_2Se_3$/$Al_2O_3$. The thickness of TiN (111) layers used in this work is around 70 nm thick. The thickness of $Bi_2Se_3$ layers in Figure 1 is 60 nm.

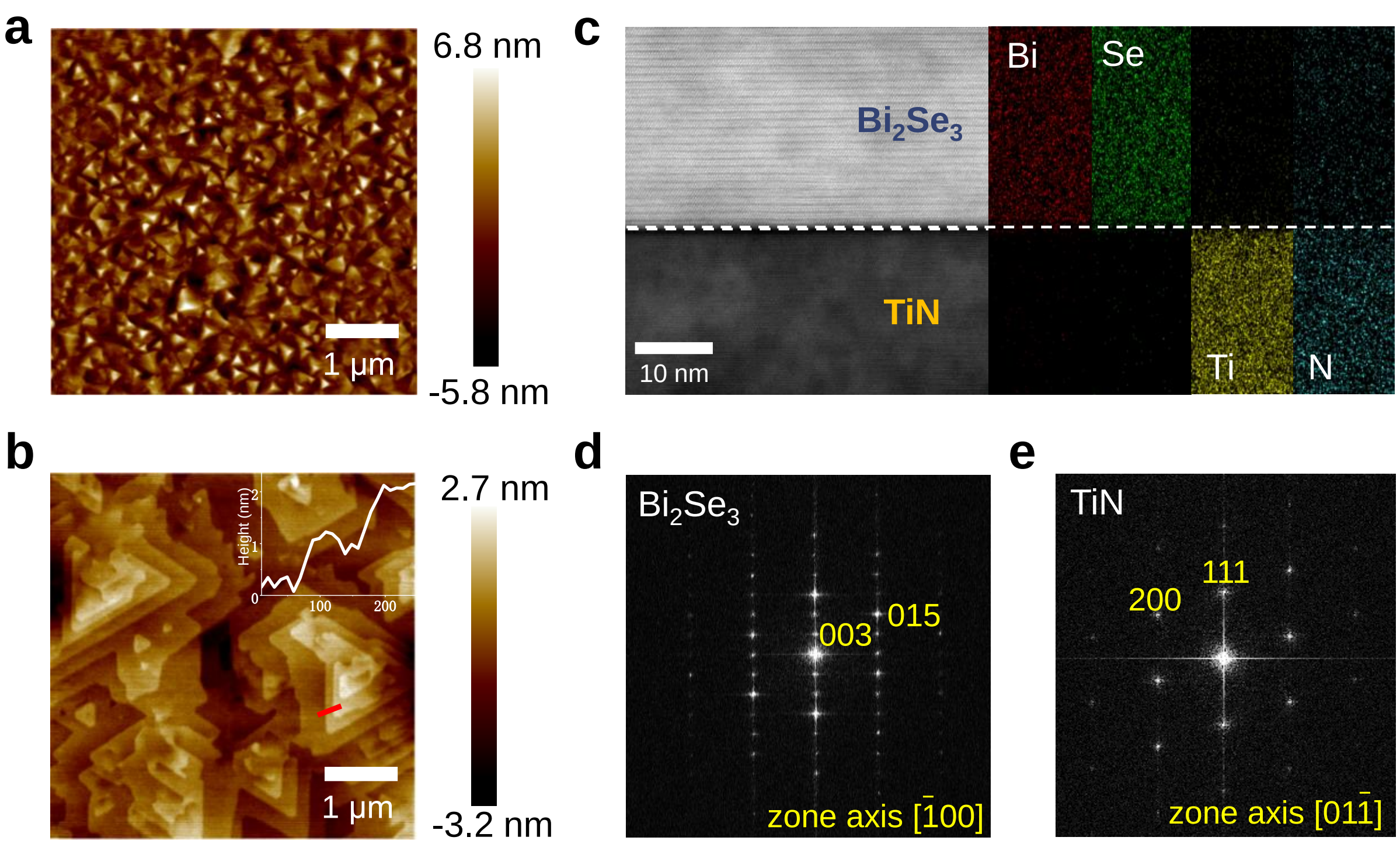

Figure 2. (a, b) The AFM morphology of (a) $Bi_2Se_3/Al_2O_3$ and (b) $Bi_2Se_3/TiN/Al_2O_3$, respectively. The inset image in (b) shows the height profile of the $Bi_2Se_3$ terrace grown on TiN and indicates the step is about 1 nm, corresponding to 1 QL of $Bi_2Se_3$. The thickness of the $Bi_2Se_3$ layer is around 60 nm in (a) and (b). (c) STEM-HAADF image of the $Bi_2Se_3$/TiN heterostructure in (b), combined with the EDX mapping results for Bi, Se, Ti and N. (d, e) The Fourier transformed image at (d) $Bi_2Se_3$ layer and (e) TiN layer.

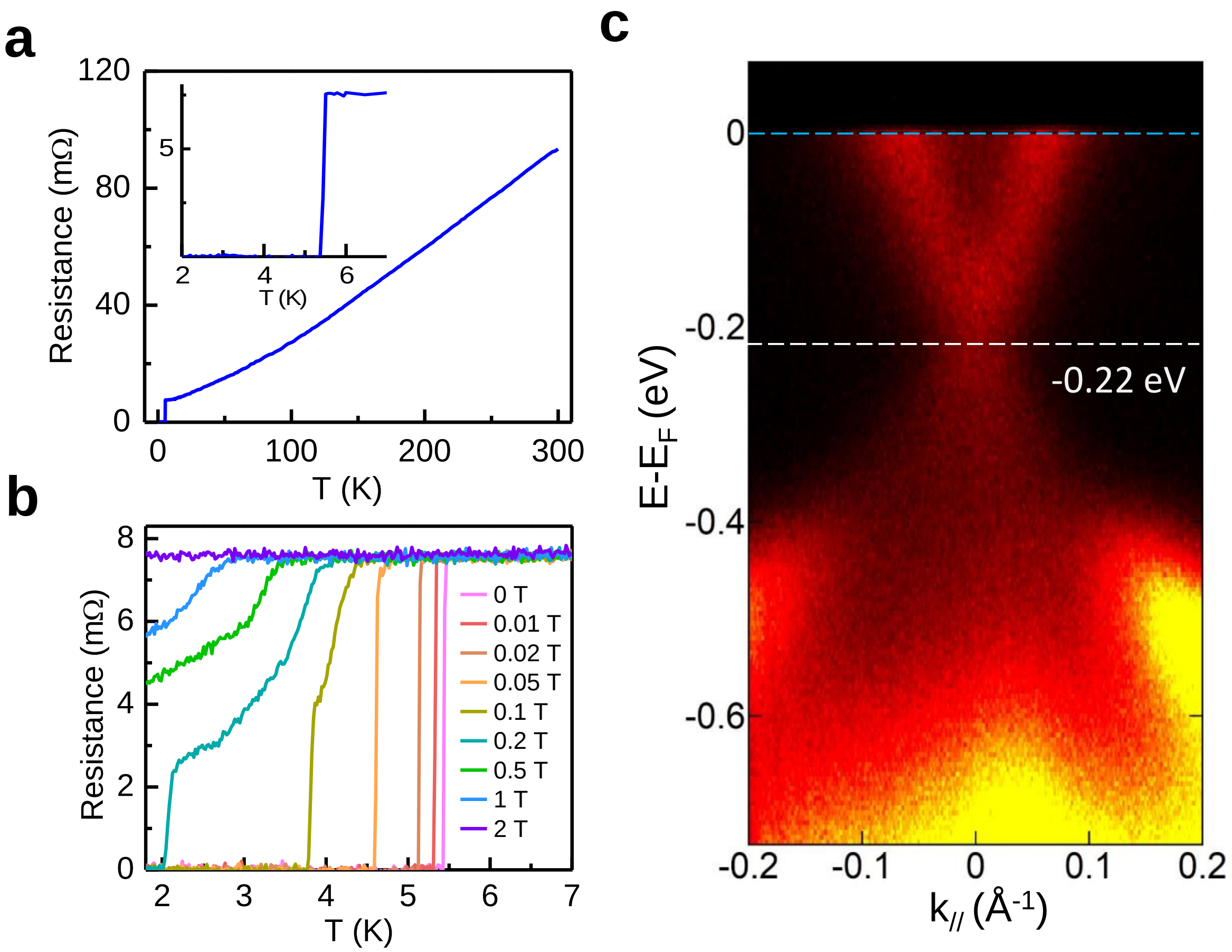

Figure 3. (a) Temperature-dependent longitudinal resistance of a $Bi_2Se_3$/TiN heterostructure from 300 K to 1.8 K. Inset shows the enlarged plot from 1.8 K to 7 K and indicates the $T_C$ is about 5.4 K. The thickness of $Bi_2Se_3$ layer is 60 nm. (b) Temperature-dependent longitudinal resistance of the same sample measured under varying magnetic fields perpendicular to the *ab* plane. (c) The band structure of a 15 nm-thick $Bi_2Se_3$ film grown on TiN characterized by ARPES with a 177 nm deep-ultraviolet laser. The data was collected at 7.3 K. The horizontal dash lines show the crossing of the Fermi level ($E-E_F = 0$) and the Dirac crossing.

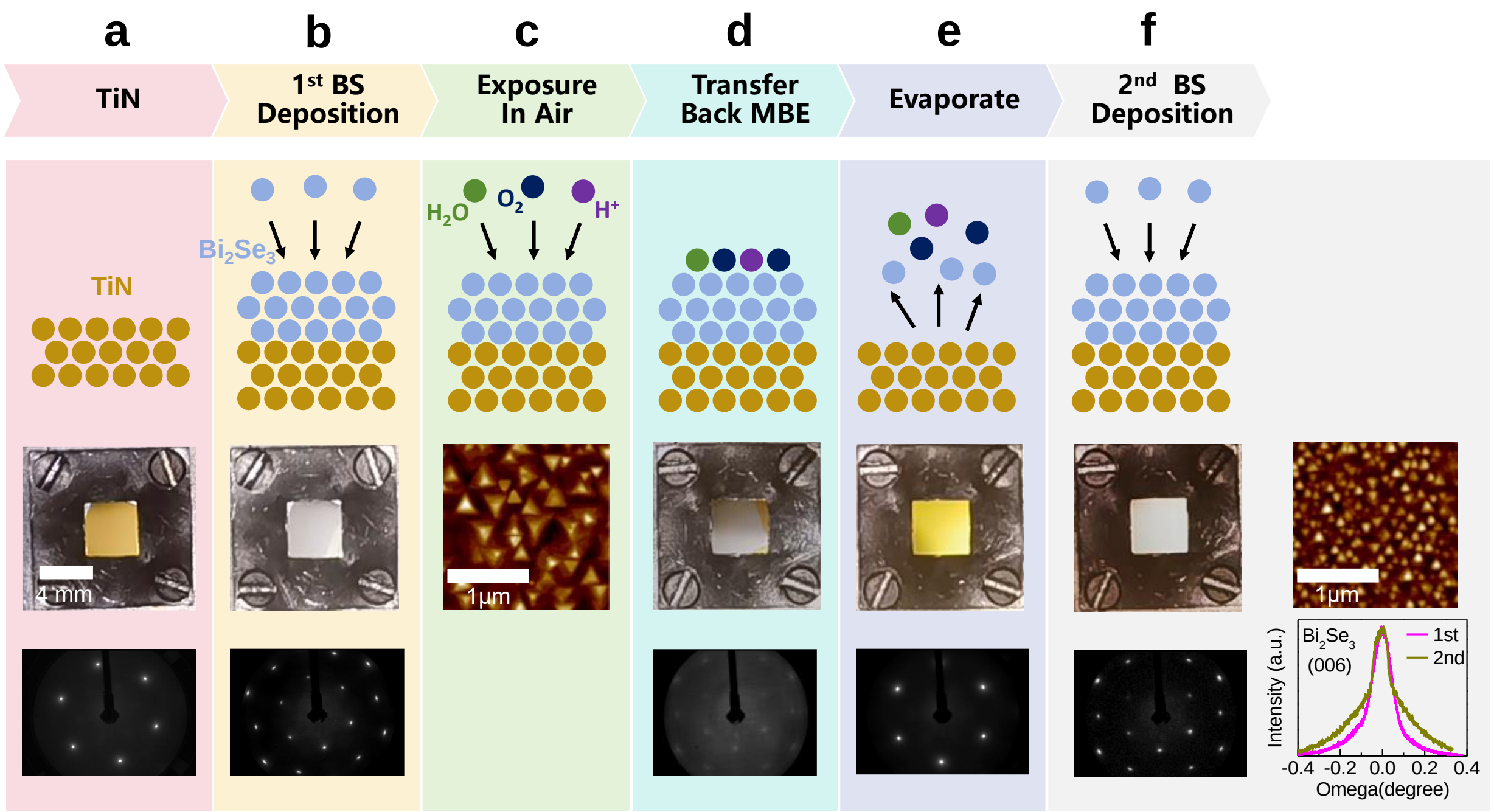

Figure 4. Schematic illustration of the recyclable growth of $Bi_2Se_3$/TiN heterostructure. Cartoons show each stage of deposition or evaporation of the film. All the pictures and LEED patterns are collected in situ during each step. The AFM images were taken in air exposure and the scale is 2μm × 2μm. (a-c) The first time $Bi_2Se_3$ deposition and AFM characterization in air exposure. (d, e) The evaporation of the $Bi_2Se_3$ layer together with pollutions from air exposure. (f) The second time $Bi_2Se_3$ deposition together with the corresponding AFM image and rocking curves.

Supporting information:

The surface flatness of $Al_2O_3$ substrate and TiN film, in-plane XRD φ scans, Thickness measurement by AFM, STEM and XRR, the XRD profile of the second time grown $Bi_2Se_3$/TiN, the transport property data of TiN after $Bi_2Se_3$ evaporation

AUTHOR INFORMATION

Corresponding Authors

*E-mail: ywcao@nimte.ac.cn

*E-mail: yaoxiong@nimte.ac.cn

Author Contributions

R.X. and M.G. contribute equally to this work. X.Y. and Y.C. conceived the experiments. R.X., X.Y., H.Y., S.O., and X.Y. grew the $Bi_2Se_3$/TiN heterostructures. J.Z., J.B., and Y.C. grew the TiN films. M.G. performed the STEM measurements and analyzed the data. R.X., B.W., and X.Y. performed the transport measurements. R.X. and S.X. performed the ARPES measurements. X.Y. and R.X. analyzed the results and wrote the manuscript with contributions from all authors.

Notes

The authors declare no competing financial interest.

ACKNOWLEDGMENTS

The work is supported by the National Natural Science Foundation of China (Grant Nos. 12304541, 12204495, U2032126, and U2032207), the National Key R&D Program of China (Grant No. 2022YFA1403000), the Zhejiang Provincial Natural Science Foundation of China (Grant No. LD24F040001), and the Ningbo Science and Technology Bureau (Grant Nos. 2023J047 and 2022Z086). The work at Rutgers University is supported by Army Research Office's W911NF2010108 and MURI W911NF2020166.

**REFERENCES**

1.Fu, L.; Kane, C. L. Superconducting proximity effect and majorana fermions at the surface of a topological insulator. Phys. Rev. Lett. **2008,** 100 (9), 096407.

2.Sato, M.; Ando, Y. Topological superconductors: a review. Rep. Prog. Phys. **2017,** 80 (7), 076501.

3.Xu, J. P.; Wang, M. X.; Liu, Z. L.; Ge, J. F.; Yang, X.; Liu, C.; Xu, Z. A.; Guan, D.; Gao, C. L.; Qian, D.; Liu, Y.; Wang, Q. H.; Zhang, F. C.; Xue, Q. K.; Jia, J. F. Experimental detection of a Majorana mode in the core of a magnetic vortex inside a topological insulator-superconductor $Bi_2Te_3/NbSe_2$ heterostructure. Phys. Rev. Lett. **2015,** 114 (1), 017001.

4.Sun, H. H.; Zhang, K. W.; Hu, L. H.; Li, C.; Wang, G. Y.; Ma, H. Y.; Xu, Z. A.; Gao, C. L.; Guan, D. D.; Li, Y. Y.; Liu, C.; Qian, D.; Zhou, Y.; Fu, L.; Li, S. C.; Zhang, F. C.; Jia, J. F. Majorana zero mode detected with spin selective Andreev reflection in the vortex of a topological superconductor. Phys. Rev. Lett. **2016,** 116 (25), 257003.

5.Qi, X.-L.; Hughes, T. L.; Zhang, S.-C. Chiral topological superconductor from the quantum Hall state. Phys. Rev. B **2010,** 82 (18), 184516.

6.Chung, S. B.; Qi, X.-L.; Maciejko, J.; Zhang, S.-C. Conductance and noise signatures of Majorana backscattering. Phys. Rev. B **2011,** 83 (10), 100512.

7.He, Q. L.; Pan, L.; Stern, A. L.; Burks, E. C.; Che, X.; Yin, G.; Wang, J.; Lian, B.; Zhou, Q.; Choi, E. S.; Murata, K.; Kou, X.; Chen, Z.; Nie, T.; Shao, Q.; Fan, Y.; Zhang, S. C.; Liu, K.; Xia,

J.; Wang, K. L. Chiral Majorana fermion modes in a quantum anomalous Hall insulator-superconductor structure. Science **2017,** 357 (6348), 294-299.

8.Kayyalha, M.; Xiao, D.; Zhang, R.; Shin, J.; Jiang, J.; Wang, F.; Zhao, Y. F.; Xiao, R.; Zhang, L.; Fijalkowski, K. M.; Mandal, P.; Winnerlein, M.; Gould, C.; Li, Q.; Molenkamp, L. W.; Chan, M. H. W.; Samarth, N.; Chang, C. Z. Absence of evidence for chiral Majorana modes in quantum anomalous Hall-superconductor devices. Science **2020,** 367 (6473), 64-67.

9.Wang, E.; Ding, H.; Fedorov, A. V.; Yao, W.; Li, Z.; Lv, Y.-F.; Zhao, K.; Zhang, L.-G.; Xu, Z.; Schneeloch, J.; Zhong, R.; Ji, S.-H.; Wang, L.; He, K.; Ma, X.; Gu, G.; Yao, H.; Xue, Q.-K.; Chen, X.; Zhou, S. Fully gapped topological surface states in $Bi_2Se_3$ films induced by a d-wave high-temperature superconductor. Nat. Phys. **2013,** 9 (10), 621-625.

10.Flototto, D.; Ota, Y.; Bai, Y.; Zhang, C.; Okazaki, K.; Tsuzuki, A.; Hashimoto, T.; Eckstein, J. N.; Shin, S.; Chiang, T. C. Superconducting pairing of topological surface states in bismuth selenide films on niobium. Sci. Adv. **2018,** 4 (4), eaar7214.

11.Wang, M. X.; Liu, C.; Xu, J. P.; Yang, F.; Miao, L.; Yao, M. Y.; Gao, C. L.; Shen, C.; Ma, X.; Chen, X.; Xu, Z. A.; Liu, Y.; Zhang, S. C.; Qian, D.; Jia, J. F.; Xue, Q. K. The coexistence of superconductivity and topological order in the $Bi_2Se_3$ thin films. Science **2012,** 336 (6077), 52-55.

12.Yang, H.; Li, Y. Y.; Liu, T. T.; Xue, H. Y.; Guan, D. D.; Wang, S. Y.; Zheng, H.; Liu, C. H.; Fu, L.; Jia, J. F. Superconductivity of topological surface states and strong proximity effect in $Sn_{1-x}Pb_xTe$-Pb Heterostructures. Adv. Mater. **2019,** 31 (52), e1905582.

13.Hlevyack, J. A.; Najafzadeh, S.; Lin, M.-K.; Hashimoto, T.; Nagashima, T.; Tsuzuki, A.; Fukushima, A.; Bareille, C.; Bai, Y.; Chen, P.; Liu, R.-Y.; Li, Y.; Flötotto, D.; Avila, J.; Eckstein, J. N.; Shin, S.; Okazaki, K.; Chiang, T. C. Massive suppression of proximity pairing in topological $(Bi_{1-x}Sb_x)_2Te_3$ films on niobium. Phys. Rev. Lett. **2020,** 124 (23), 236402.

14.Trang, C. X.; Shimamura, N.; Nakayama, K.; Souma, S.; Sugawara, K.; Watanabe, I.; Yamauchi, K.; Oguchi, T.; Segawa, K.; Takahashi, T.; Ando, Y.; Sato, T. Conversion of a conventional superconductor into a topological superconductor by topological proximity effect. Nat. Commun. **2020,** 11 (1), 159.

15.Yi, H.; Hu, L. H.; Wang, Y.; Xiao, R.; Cai, J.; Hickey, D. R.; Dong, C.; Zhao, Y. F.; Zhou, L. J.; Zhang, R.; Richardella, A. R.; Alem, N.; Robinson, J. A.; Chan, M. H. W.; Xu, X.; Samarth, N.; Liu, C. X.; Chang, C. Z. Crossover from Ising- to Rashba-type superconductivity in epitaxial $Bi_2Se_3$/monolayer $NbSe_2$ heterostructures. Nat. Mater. **2022,** 21 (12), 1366-1372.

16.Yi, H.; Hu, L. H.; Zhao, Y. F.; Zhou, L. J.; Yan, Z. J.; Zhang, R.; Yuan, W.; Wang, Z.; Wang, K.; Hickey, D. R.; Richardella, A. R.; Singleton, J.; Winter, L. E.; Wu, X.; Chan, M. H. W.; Samarth, N.; Liu, C. X.; Chang, C. Z. Dirac-fermion-assisted interfacial superconductivity in epitaxial topological-insulator/iron-chalcogenide heterostructures. Nat. Commun. **2023,** 14 (1), 7119.

17.Yi, H.; Zhao, Y. F.; Chan, Y. T.; Cai, J.; Mei, R.; Wu, X.; Yan, Z. J.; Zhou, L. J.; Zhang, R.; Wang, Z.; Paolini, S.; Xiao, R.; Wang, K.; Richardella, A. R.; Singleton, J.; Winter, L. E.; Prokscha, T.; Salman, Z.; Suter, A.; Balakrishnan, P. P.; Grutter, A. J.; Chan, M. H. W.; Samarth, N.; Xu, X.; Wu, W.; Liu, C. X.; Chang, C. Z. Interface-induced superconductivity in magnetic topological insulators. Science **2024,** 383 (6683), 634-639.

18.Moore, R. G.; Lu, Q.; Jeon, H.; Yao, X.; Smith, T.; Pai, Y. Y.; Chilcote, M.; Miao, H.; Okamoto, S.; Li, A. P.; Oh, S.; Brahlek, M. Monolayer superconductivity and tunable topological electronic structure at the Fe(Te,Se)/$Bi_2Te_3$ interface. Adv. Mater. **2023,** 35 (22), e2210940.

19.Liang, J.; Zhang, Y. J.; Yao, X.; Li, H.; Li, Z. X.; Wang, J.; Chen, Y.; Sou, I. K. Studies on the origin of the interfacial superconductivity of $Sb_2Te_3/Fe_{1+y}Te$ heterostructures. Proc. Natl. Acad. Sci. U.S.A. **2020,** 117 (1), 221-227.

20.Yao, X.; Brahlek, M.; Yi, H. T.; Jain, D.; Mazza, A. R.; Han, M. G.; Oh, S. Hybrid symmetry epitaxy of the superconducting Fe(Te,Se) film on a topological insulator. Nano Lett. **2021,** 21 (15), 6518-6524.

21.Yao, X.; Gao, B.; Han, M. G.; Jain, D.; Moon, J.; Kim, J. W.; Zhu, Y.; Cheong, S. W.; Oh, S. Record high-proximity-induced anomalous Hall effect in $(Bi_xSb_{1-x})_2Te_3$ thin film grown on $CrGeTe_3$ substrate. Nano Lett. **2019,** 19 (7), 4567-4573.

22.Yao, X.; Moon, J.; Cheong, S.-W.; Oh, S. Structurally and chemically compatible $BiInSe_3$ substrate for topological insulator thin films. Nano Res. **2020,** 13 (9), 2541-2545.

23.Yao, X.; Yi, H. T.; Jain, D.; Oh, S. Suppressing carrier density in $(Bi_xSb_{1-x})_2Te_3$ films using $Cr_2O_3$ interfacial layers. J. Phys. D: Appl. Phys. **2021,** 54 (50), 504007.

24.Yao, X.; Mazza, A. R.; Han, M. G.; Yi, H. T.; Jain, D.; Brahlek, M.; Oh, S. Superconducting fourfold Fe(Te,Se) film on sixfold magnetic MnTe via hybrid symmetry epitaxy. Nano Lett. **2022,** 22 (18), 7522-7526.

25.Yilmaz, T.; Pletikosić, I.; Weber, A. P.; Sadowski, J. T.; Gu, G. D.; Caruso, A. N.; Sinkovic, B.; Valla, T. Absence of a proximity effect for a thin-films of a $Bi_2Se_3$ topological insulator grown on top of a $Bi_2Sr_2CaCu_2O_{8+\delta}$ cuprate superconductor. Phys. Rev. Lett. **2014,** 113 (6), 067003.

26.Wan, S.; Gu, Q.; Li, H.; Yang, H.; Schneeloch, J.; Zhong, R. D.; Gu, G. D.; Wen, H.-H. Twofold symmetry of proximity-induced superconductivity in $Bi_2Te_3/Bi_2Sr_2CaCu_2O_{8+\delta}$ heterostructures revealed by scanning tunneling microscopy. Phys. Rev. B **2020,** 101 (22), 220503.

27.Zhang, H. J.; Liu, C. X.; Qi, X. L.; Dai, X.; Fang, Z.; Zhang, S. C. Topological insulators in $Bi_2Se_3$, $Bi_2Te_3$ and $Sb_2Te_3$ with a single Dirac cone on the surface. Nat. Phys. **2009,** 5 (6), 438-442.

28.Zhang, W.; Yu, R.; Zhang, H.-J.; Dai, X.; Fang, Z. First-principles studies of the three-dimensional strong topological insulators $Bi_2Te_3$, $Bi_2Se_3$ and $Sb_2Te_3$. New Journal of Physics **2010,** 12 (6), 065013.

29.Bansal, N.; Kim, Y. S.; Edrey, E.; Brahlek, M.; Horibe, Y.; Iida, K.; Tanimura, M.; Li, G.-H.; Feng, T.; Lee, H.-D.; Gustafsson, T.; Andrei, E.; Oh, S. Epitaxial growth of topological insulator $Bi_2Se_3$ film on Si(111) with atomically sharp interface. Thin Solid Films **2011,** 520 (1), 224-229.

30.Brahlek, M.; Koirala, N.; Salehi, M.; Moon, J.; Zhang, W.; Li, H.; Zhou, X.; Han, M.-G.; Wu, L.; Emge, T.; Lee, H.-D.; Xu, C.; Rhee, S. J.; Gustafsson, T.; Armitage, N. P.; Zhu, Y.; Dessau, D. S.; Wu, W.; Oh, S. Disorder-driven topological phase transition in $Bi_2Se_3$ films. Phys. Rev. B **2016,** 94 (16), 165104.

31.Koirala, N.; Brahlek, M.; Salehi, M.; Wu, L.; Dai, J.; Waugh, J.; Nummy, T.; Han, M. G.; Moon, J.; Zhu, Y.; Dessau, D.; Wu, W.; Armitage, N. P.; Oh, S. Record surface state mobility and quantum Hall effect in topological insulator thin films via interface engineering. Nano Lett. **2015,** 15 (12), 8245-8249.

32.Hegde, S. S.; Yue, G.; Wang, Y.; Huemiller, E.; Van Harlingen, D. J.; Vishveshwara, S. A topological Josephson junction platform for creating, manipulating, and braiding Majorana bound states. Annals of Physics **2020,** 423, 168326.

33.Abboud, N.; Subramanyan, V.; Sun, X.-Q.; Yue, G.; Van Harlingen, D.; Vishveshwara, S. Signatures of Majorana bound states and parity effects in two-dimensional chiral p-wave Josephson junctions. Phys. Rev. B **2022,** 105 (21), 214521.

34.Yue, G.; Zhang, C.; Huemiller, E. D.; Montone, J. H.; Arias, G. R.; Wild, D. G.; Zhang, J. Y.; Hamilton, D. R.; Yuan, X.; Yao, X.; Jain, D.; Moon, J.; Salehi, M.; Koirala, N.; Oh, S.; Van Harlingen, D. J. Signatures of Majorana bound states in the diffraction patterns of extended superconductor–topological insulator–superconductor Josephson junctions. Phys. Rev. B **2024,** 109 (9), 094511.

35.Zhang, R.; Li, X.; Meng, F.; Bi, J.; Zhang, S.; Peng, S.; Sun, J.; Wang, X.; Wu, L.; Duan, J.; Cao, H.; Zhang, Q.; Gu, L.; Huang, L. F.; Cao, Y. Wafer-scale epitaxy of flexible nitride films with superior plasmonic and superconducting performance. ACS Appl Mater Interfaces **2021,** 13 (50), 60182-60191.

36.Zhang, R.; Ma, Q.-Y.; Liu, H.; Sun, T.-Y.; Bi, J.; Song, Y.; Peng, S.; Liang, L.; Gao, J.; Cao, H.; Huang, L.-F.; Cao, Y. Crystal orientation-dependent oxidation of epitaxial TiN films with tunable plasmonics. ACS Photonics **2021,** 8 (3), 847-856.

37.Zhang, R.; Lin, T.; Peng, S.; Bi, J.; Zhang, S.; Su, G.; Sun, J.; Gao, J.; Cao, H.; Zhang, Q.; Gu, L.; Cao, Y. Flexible but refractory single-crystalline hyperbolic metamaterials. Nano Lett. **2023,** 23 (9), 3879-3886.

38.Bi, J.; Lin, Y.; Zhang, Q.; Liu, Z.; Zhang, Z.; Zhang, R.; Yao, X.; Chen, G.; Liu, H.; Huang, Y.; Sun, Y.; Zhang, H.; Sun, Z.; Xiao, S.; Cao, Y. Momentum-resolved electronic structures and strong electronic correlations in graphene-like nitride superconductors. Nano Lett. **2024**.